\documentclass[journal=nalefd,manuscript=letter]{achemso}

\usepackage[version=3]{mhchem} % Formula subscripts using \ce{}
\usepackage{eurosym}
\usepackage{mathtools}

\newcommand{\ud}[1]{#1^{\dag}} 
\newcommand{\bra}[1]{\left\langle #1\right|}
\newcommand{\ket}[1]{\left| #1\right\rangle}

\author{Fabrice P.~Laussy} 
\affiliation[Walter Schottky Institut and Physikdepartment, Technische Universit\"at M\"unchen,
Am Coulombwall 4, 85748 Garching, Germany]{Walter Schottky Institut (M\"unchen)}
\alsoaffiliation[F\'isica Te\'orica de la Materia Condensada,
  Universidad Aut\'onoma de Madrid, 28049, Madrid, Spain]{Universidad Aut\'onoma de Madrid}
\email{fabrice.laussy@gmail.com}

\author{Vase Jovanov}
\affiliation[Walter Schottky Institut and Physikdepartment, Technische Universit\"at M\"unchen,
Am Coulombwall 4, 85748 Garching, Germany]{Walter Schottky Institut (M\"unchen)}

\author{Elena del Valle}
\affiliation[Physikdepartment, Technische Universit\"at M\"unchen, 
James Franck Str., 85748 Garching, Germany]{Technische Universit\"at M\"unchen}

\author{Alexander Bechtold}
\author{Stephan Kapfinger}
\author{Kai~M\"uller}
\author{Sebastian Koch}
\author{Arne Laucht}
\author{Thomas~Eissfeller}
\author{Max Bichler}
\author{Gerhard Abstreiter}
\author{Jonathan J. Finley}
\affiliation[Walter Schottky Institut and Physikdepartment, Technische Universit\"at M\"unchen,
Am Coulombwall 4, 85748 Garching, Germany]{Walter Schottky Institut (M\"unchen)}

\title[Fluctuation induced luminescence]{Fluctuation induced
  luminescence sidebands in the emission spectra of resonantly driven
  quantum dots}

\abbreviations{QD,QDM,PL} 


\begin{document}
%%%%%%%%%%%%%%%%%%%%%%%%%%%%%%%%%%%%%%%%%%%%%%%%%%%%%%%%%%%%%%%%%%%%%
%% The manuscript does not need to include \maketitle, which is
%% executed automatically.  The document should begin with an
%% abstract, if appropriate.  If one is given and should not be, the
%% contents will be gobbled.
%%%%%%%%%%%%%%%%%%%%%%%%%%%%%%%%%%%%%%%%%%%%%%%%%%%%%%%%%%%%%%%%%%%%%
\begin{abstract}
  We describe how complex fluctuations of the local environment of an
  optically active quantum dot can leave rich fingerprints in its
  emission spectrum.  A new feature, termed ``Fluctuation Induced
  Luminescence'' (FIL), is observed to arise from extremely rare
  fluctuation events that have a dramatic impact on the response of
  the system---so called ``black swan'' events. A quantum dissipative
  master equation formalism is developed to describe this effect
  phenomenologically. Experiments performed on single quantum dots
  subject to electrical noise show excellent agreement with our
  theory, producing the characteristic FIL sidebands.
\end{abstract}

%\section{Introduction}

Discrete solid-state quantum emitters such as III-V quantum
dots\cite{wang_book08a} and NV-centers\cite{awschalom07a} represent
highly versatile hardware for all-optical quantum technologies. In
particular, self-assembled InGaAs Quantum Dots (QDs) are robust, ultra
narrow linewidth sources of triggered single photons and
polarization-entangled photon pairs.\cite{salter10a} The creation of
remote entanglement via photon-interference\cite{salter10a} calls for
discrete emitters that produce Fourier-transform limited single photon
wave packets. This is very difficult to achieve in solid-state systems
since the system typically undergoes frequency jitter due to
fluctuations of the charge environment of the
system\cite{fry00a,vamivakas11a,houel12a} and unavoidable coupling to
the lattice.\cite{muljarov05a,ramsay10b} However, for low temperatures
and weak optical driving, coupling to the lattice is
inhibited\cite{langbein04a} and the spectral lineshape is expected to
reflect slow fluctuations of the environment which manifest themselves
as discrete spectral jumps and wandering\cite{robinson00a}. Fourier
transform-limited lines are typically not achieved in optical
experiments, with measured linewidths somewhat above the theoretical
limit\cite{hogele04a,atature06a,xu07a,vamivakas09a}. While spectral
fluctuations in self-assembled QDs have been investigated with
non-resonant excitation\cite{berthelot06a,latta09a}, the impact of a
fluctuating environment in the case of true resonant excitation has
received significantly less attention. Spectral fluctuations arising
from environmental fluctuations are a common feature in condensed
matter systems arising also for colour centers in
diamond\cite{robledo10a}, colloidal nanocrystals\cite{muller04a} and
semiconductor nanowire quantum dots\cite{sallen10a}.

\begin{figure}[th]
  \includegraphics[width=\linewidth]{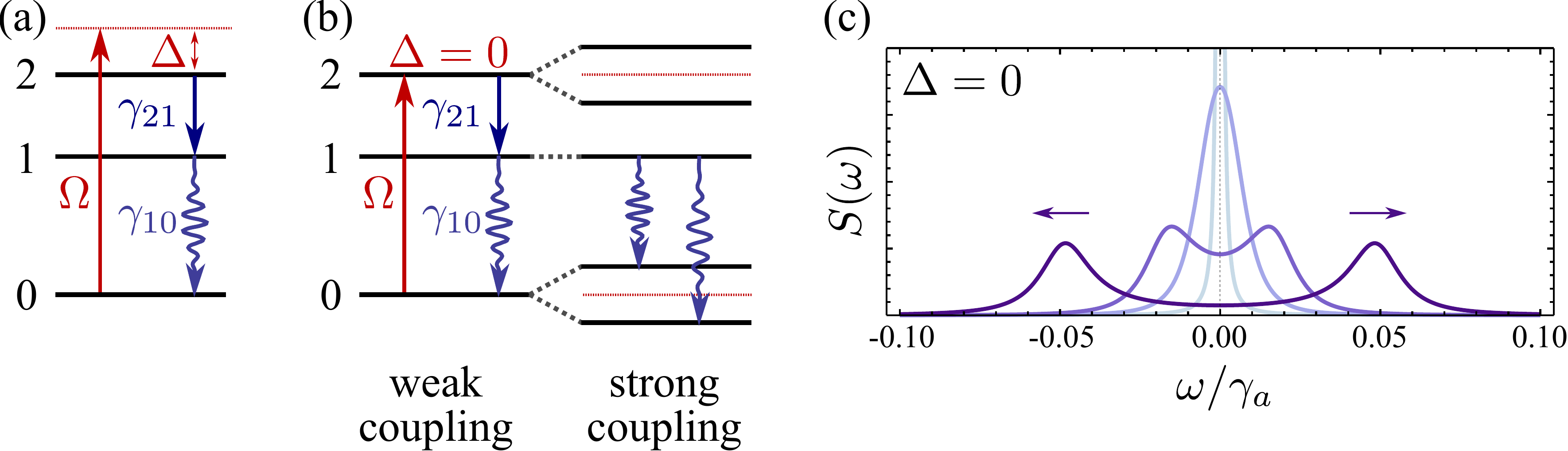}
  \caption{(a) The level structure under study, (b) dressing of the
    p-state at resonance and (c) the Mollow doublet which arises
    from it. The laser is tuned closed to the excited p orbital,
    labelled~2. Incoherent relaxation $\gamma_{21}$ transfers the
    excitation to the s-state, labelled~1, which PL is
    detected. When the excitation rate $\Omega$ is large enough, the
    laser dresses the states which are driven, namely, the p-state
    and the ground state. The latter, being common with the s-state,
    result in the observation of a Mollow doublet with splitting
    increasing linearly with $\Omega$.}
  \label{fig:FriJul20131153CEST2012}
\end{figure}

Here, we probe the impact of environmental fluctuations on the optical
response of a resonantly driven quantum emitter embedded within an
electrically tunable solid-state environment.\cite{flagg09a,muller07a}
The resonant excitation simplifies the picture since it avoids
complications due to incoherently injected carriers, such as
fluctuations arising from charge (de)trapping at defects in the
wetting later. It also allows to consider only a single projection of
spin by using circularly polarized excitation.  As depicted
schematically in \ref{fig:FriJul20131153CEST2012}(a), we consider a
situation where a narrowband single frequency laser coherently excites
the interband p-transition in a single dot and the emission from the
s-shell is monitored.\cite{ester07a} The corresponding Hamiltonian
reads:
\begin{equation}
  \label{eq:ThuJul19124257CEST2012}
    H=(\Delta-E_{21})\ket{1}\bra{1}+\Delta\ket{2}\bra{2}+\Omega (\ket{0}\bra{2}+\ket{2}\bra{0})
\end{equation}
where $\Delta=\omega_2-\omega_\mathrm{L}$ is the detuning between the
state 2 (interband p-transition) and the laser, and
$E_{21}=\omega_2-\omega_1$ the energy gap between the p and
s-interband transitions (we set $\hbar=1$ for convenience). The
relaxation between the p and s-shells---due to, e.g., phonon mediated
processes\cite{heitz01a}---occurs at the rate $\gamma_{21}$ and is
included as an incoherent relaxation in the Lindblad form,
$\frac{\gamma_{21}}2\mathcal{L}_{\ket{1}\bra{2}}$ where, for any
operator~$c$, $\mathcal{L}_c$ is the Liouville superoperator defined
as $\mathcal{L}_c(\rho)=2c\rho\ud{c}-\ud{c}c\rho-\ud{c}c\rho$. The
radiative decay of the s-state giving rise to luminescence is also
described as $\frac{\gamma_{10}}2\mathcal{L}_{\sigma}$ where
$\sigma=\ket{0}\bra{1}$. The photoluminescence (PL)
spectrum~$S(\omega, \Delta)$ is obtained by calculating
$G^{(1)}(\tau)=\langle\ud{\sigma}(0)\sigma(\tau)\rangle$ from the
master equation
$\partial_t\rho=i[\rho,H]+\frac{\gamma_{21}}2\mathcal{L}_{\ket{1}\bra{2}}(\rho)+\frac{\gamma_{10}}2\mathcal{L}_{\sigma}(\rho)$
and taking the Fourier transform. Setting as the reference energy, the
energy at which the s-state emits when $\Delta=0$ (with maximum gain),
the result reads:
%
%\begin{multline}
%  \label{eq:FriNov4121905CET2011}
%  S(\omega,\Delta)=\frac{1}{\pi}\Big[{8\gamma_{21}\Omega^2\Big(\gamma_{10}(\Gamma^2+4\omega^2)+4\Omega^2\Gamma\Big)}\Big]\Big/\Big[\Big(\gamma_{10}(\gamma_{21}^2+4\Delta^2)+4\Omega^2(2\gamma_{10}+\gamma_{21})\Big)\\\times
%  \Big((\gamma_{10}^2+4(\omega-\Delta)^2)(\Gamma^2+4
%  \omega^2)+8\Omega^2(\gamma_{10}\Gamma-4(\omega-\Delta)\omega)+16\Omega^4\Big)\Big]
%\end{multline}
%
\begin{equation}
  \label{eq:FriNov4121905CET2011}
  S(\omega,\Delta)=\frac{1}{\pi}\frac{\frac{8\gamma_{21}\Omega^2}{\gamma_{10}(\gamma_{21}^2+4\Delta^2)+4\Omega^2(2\gamma_{10}+\gamma_{21})}\Big(\gamma_{10}(\Gamma^2+4\omega^2)+4\Omega^2\Gamma\Big)}{(\gamma_{10}^2+4(\omega-\Delta)^2)(\Gamma^2+4
  \omega^2)+8\Omega^2(\gamma_{10}\Gamma-4(\omega-\Delta)\omega)+16\Omega^4}
\end{equation}
where we introduced $\Gamma\equiv\gamma_{10}+\gamma_{21}$. 

This is the full quantum picture which describes saturation of the
two-level system and dressing of the p-state by the laser when the
pumping~$\Omega$ is large.\cite{cohentannoudji77a} In this case, this
results in a spectral doublet rather than the Mollow triplet for a
two-level system,\cite{mollow69a} because relaxation is from the naked
s-state to the dressed ground state, which is common to both the bare
s and driven p levels, as shown
in~\ref{fig:FriJul20131153CEST2012}(b). Provided that the gain is
large enough, this configuration might have advantages as compared to
resonance fluorescence, since excitation and detection are detuned and
pump photons can be filtered from the emission.

\begin{figure}[th]
  \includegraphics[width=\linewidth]{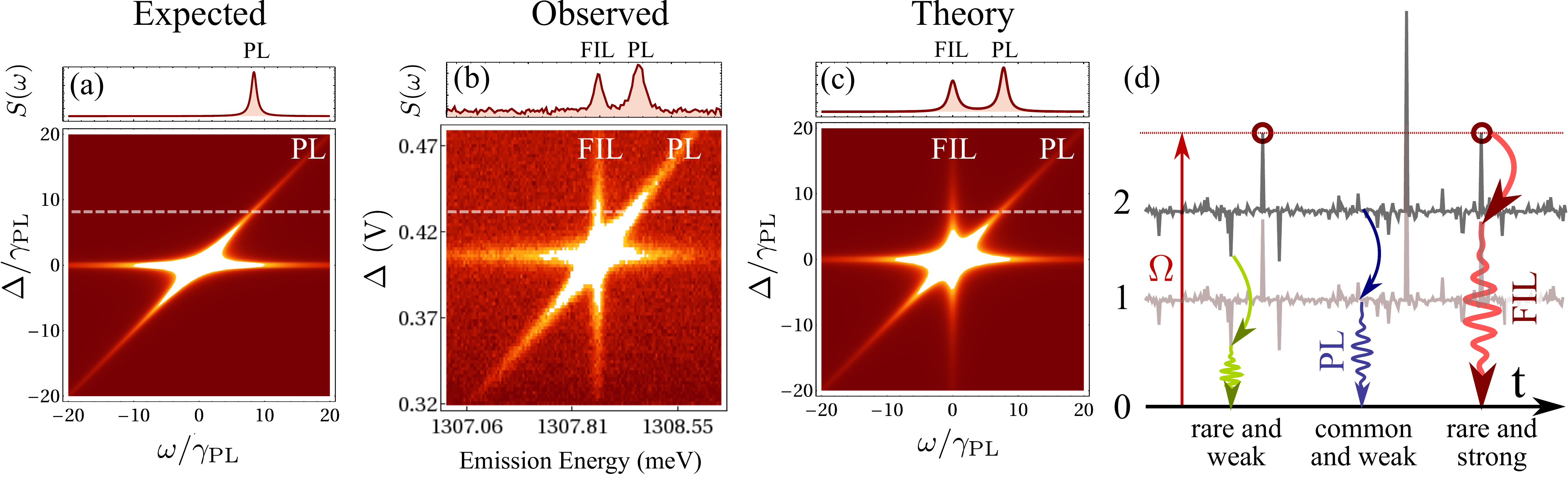}
  \caption{(a) The expected luminescence from a resonantly driven QD
    in the linear regime: the diagonal transition shows the radiative
    recombination of the s shell as the frequency of the driving
    field is detuned from the excited state, while the horizontal line
    reflects the high increase of the excitation efficiency of the
    driven oscillator.  (b) A new peak---FIL---appears in the
    experimental observation as a vertical line pinned by the laser.
    (c) The theory supplemented with fluctuations reproduces the
    observed phenomenology.  (d) Scheme of the processes which give
    rise to FIL. The levels 1 and~2 have scalefree fluctuations which,
    when they bring the system into resonance, result in a strong
    absorption and subsequent emission.}
  \label{fig:FriJul20142005CEST2012}
\end{figure}

In the linear regime, Eq.~(\ref{eq:FriNov4121905CET2011}) assumes
the much simpler form:
\begin{equation}
  \label{eq:FriNov4124436CET2011}
  S(\omega,\Delta)=
  \frac{2\Omega^2\pi}{\gamma_{10}}\times
  \frac{1}{\pi}\frac{\gamma_{21}/2}{(\gamma_{21}/2)^2+\Delta^2}\times
  \frac{1}{\pi}\frac{\gamma_{10}/2}{(\gamma_{10}/2)^2+(\omega-\Delta)^2}+O(\Omega)^4\,,
\end{equation}
with $O(\Omega^4)$ terms of at least fourth order in $\Omega$.  This
shows that the coherently excited QD luminescence is given simply by
the product of $i)$ the emitted intensity, $ii)$ the population gain
of the driven detuned oscillator, and $iii)$ the Lorentzian PL
lineshape of emission; an extremely simple and fundamental result.
The emitted intensity grows linearly with $\Omega^2$ as long as the
system remains in the linear regime under its effective
pumping~$P_\mathrm{eff}$, that is, as long as
$\mathrm{Gain}(\Delta)=P_\mathrm{eff}/\gamma_{10} \ll 1$.  At
resonance $\Delta=0$, where the driving is most efficient, the
condition reads
$\mathrm{Gain}(0)=4\Omega^2/(\gamma_{21}\gamma_{10})\ll 1$. We will
restrict our discussion to the linear regime in the remainder of the
manuscript, where one expects only emission from the s-shell directly
to the ground state by direct radiative recombination.  Plotting the
intensity of emission of the s-state photons (in colour, with lighter
shades corresponding to higher intensities) detected at a given energy
(on the $x$ axis) for the various detunings of the lasers with the
p-state (on the $y$ axis), this produces a tilted line, since the
emission linearly tracks the energy detuning from resonance. This
feature is presented by the diagonally shifting transition shown
in~\ref{fig:FriJul20142005CEST2012}(a). The horizontal line
superimposed at $\Delta=0$ arises from the strong enhancement of the
emission signal at resonance, termed ``gain'' in the discussion below.

Measurements were made on many different single quantum dots and
QD-Molecules (QDMs)\cite{muller11a,muller12a} to test the predictions
of this theory.  As discussed below and presented in
\ref{fig:FriJul20142005CEST2012}(b), the experimental findings differ
significantly from these simple expectations. The samples investigated
were GaAs n-i--Schottky photodiodes containing a low density layer of
InGaAs self-assembled QDs or vertically stacked QDMs. Such structures
facilitate control of the static electric field in the vicinity of the
nanostructure via an applied voltage. The growth conditions used gave
rise to QD nanostructures with a large radius (30-50nm) and, thus, a
large oscillator strength as evidenced by measurements of short
radiative lifetimes for self-assembled nanostructures ($\leq 700$ps,
limited by the temporal resolution of our setup).  Individual QDs or
QDMs were excited optically using a tunable single frequency laser and
their photoluminescence was recorded using a low temperature confocal
microscope. Photoluminescence excitation (PLE) spectra from the same
dot were obtained by recording the emission in the vicinity of s-shell
transition with a CCD camera while scanning the laser energy through
the excited orbital states of the system---fully analogous to the
theoretical scheme introduced in the discussion related to
\ref{fig:FriJul20131153CEST2012}.  A typical experimental result is
presented in \ref{fig:FriJul20142005CEST2012}(b) in a false colour
representation for excitation in the range 9-13meV above the s-shell
neutral exciton transition.  An unexpected vertical line is clearly
observed, labelled FIL in \ref{fig:FriJul20142005CEST2012}(b), that is
pinned to zero detuning. The energy gap between this newly observed
FIL feature and the s-shell emission is identical to the detuning of
the laser from the p-shell transition. In a configuration where the
single mode laser was fixed and the p and s transition were shifted,
for instance by application of an electric field to tune the levels
through the DC Stark effect, we obtained identical results.  This
observation was also found to hold for QD-molecules pumping the
antibonding branch of the neutral, coupled exciton\cite{krenner05b}
and detecting the bonding branch.  Here, the energy gap between the
two transitions involved varies with electric field\cite{muller12a}
but the FIL peak always appears at the energy of maximum gain.  We
will now show that this striking phenomenology is fully accounted for
by fluctuations, albeit of a particular type.

We proceed with the introduction of fluctuations in the theoretical
model, an undertaking that is nontrivial at a microscopic
level\cite{richter09a}. Our modelling remains rooted in the master
equation approach and hinges on a reservoir that can be in any of
$N_R+1$ configurational states, labelled as $R=-N_R/2,\hdots, 0,
\hdots,N_R/2$, each of which has the effect of shifting the levels of
our system according to $\omega_{iR}=\omega_i+R \delta $,
$i=1,2$. Such a picture has been introduced by Budini\cite{budini09a}
with a small reservoir to describe blinking. Here, we will assume that
the reservoir is large enough to approximate a continuous fluctuating
bath, introducing the need of a continuous distribution for the
fluctuations. The driven QD Hamiltonian now depends on the states of
the reservoir:
\begin{equation}
  \label{eq:MonNov21163135CET2011}
  \tilde H_R=(\Delta-E_{21}+R \delta)\ket{1R}\bra{1R}+(\Delta+R\delta)\ket{2R}\bra{2R}+\Omega (\ket{0R}\bra{2R}+\ket{2R}\bra{0R})
\end{equation}
where we use a `\textasciitilde' symbol to denote that a fluctuating
environment has been included in the picture. The states $\ket{iR}$,
with $i=0,1$ or~2, correspond to the case where the QD is in the state
$\ket{i}$ and the reservoir in its configuration~$R$. Note that the
latter is not a proper quantum number, but a label to index a
macrostate for the environment. We assume that only the energy of the
levels fluctuates while other parameters remain unchanged. In this
case, the master equation of the system now reads:
\begin{subequations}
  \label{eq:MonNov21163729CET2011}
  \begin{align}
    \partial_t\rho_R=&i[\rho_R,\tilde H_R]+\frac{\gamma_{10}}{2}\mathcal{L}_{\ket{0R}\bra{1R}}(\rho_R)+\frac{\gamma_{21}}{2}\mathcal{L}_{\ket{1R}\bra{2R}}(\rho_R)\\
    -&\sum_{R'}\phi_{R\rightarrow
      R'}\rho_R+\sum_{R'}\phi_{R'\rightarrow R}\rho_{R'}\,.
  \end{align}
\end{subequations}
We assume that the transitions between the configurational states of
the reservoir do not depend on the state of the system. As neither the
initial condition nor the dynamics introduces any coherence between
the bath macrostates, their density matrix can be written as
$\rho_{Env}=\sum_R \mathcal{P}_R \ket{R}\bra{R}$ where
$\mathcal{P}_R=\mathrm{Tr}_S(\rho_R)$ are the populations. The
fluctuations between the configurational states of the reservoir are,
therefore, ruled by a classical evolution: $\partial_t
\mathcal{P}_R=-\sum_{R'}\phi_{R\rightarrow
  R'}\mathcal{P}_R+\sum_{R'}\phi_{R'\rightarrow R}\mathcal{P}_{R'}$,
defining the kinetic dynamics of the environment. On the other hand,
the system dynamics arise after tracing out all internal transitions
between the configurational states of the reservoir,
$\rho=\sum_R\rho_R$. Therefore, the evolution of the system density
matrix is non-Markovian, $\partial_t \rho=\mathbf{L}(\rho)+\int_0^t
d\tau \mathcal{K}(t-\tau)\rho(\tau)$, where $\mathbf{L}$ represents
the unitary dynamics in the absence of the reservoir and $\mathcal{K}$
the memory kernel introduced by it. One can compute the kernel
analytically when the number of $R$-states is small. In this case, the
steady state ($t=0$) correlator is $G^{(1)}(\tau)=\sum_{R,R'} \langle
\sigma^\dagger_{R'} (0)\sigma_R(\tau)\rangle$ where
$\sigma_R=\ket{0R}\bra{1R}$. Introducing an auxiliary matrix
$\theta_R$ that follows the same master equation as the conditional
density matrix $\rho_R$ but with initial conditions (i.~c.) that
depend on a different state of the reservoir $R'$, i.e., $\langle
k|\theta_r(0)|l\rangle =\delta_{r,R'}\delta_{l,0}\langle
k|\rho_{R'}|1\rangle$, each correlator $\langle \sigma^\dagger_{R'}
(0)\sigma_R(\tau)\rangle$ can be computed in terms of the elements of
$\theta_R$ according to $\langle \sigma^\dagger_{R'} (0)
\sigma_R(\tau)\rangle=\sum_{k,l}\langle
k|\theta_R(\tau)|l\rangle\langle l| \sigma_R|k\rangle=\langle
1|\theta_R(\tau)|0\rangle$.
With this, the total correlator reads:
\begin{equation}
  \label{eq:ThuNov24111710CET2011}
  G^{(1)}(\tau)=\sum_R \Big\{ \sum_{R'} \Big[ \langle 1|\theta_R(\tau)|0\rangle\Big]_{\text{i.~c.: }\langle k|\theta_{R'}(0)|0\rangle=\langle k|\rho_{R'}|1\rangle}\Big\}\,.
\end{equation}
Its Fourier transform provides the PL spectrum $\tilde
S(\omega,\Delta)$ in the fluctuating environment. The formalism is
general and can be applied to arbitrary quantum optical systems where
fluctuations are believed to play an important
role\cite{hennessy07a,ota09b}.

Now that we have formally included the fluctuations at a microscopic
level, we need to specify their character. They have been left
completely arbitrary in the discussion until now.  Since Gauss
introduced his law of errors\cite{gauss_book1809a}, the so-called
``normal distribution'' became the archetypal fluctuation in physical
systems. Its importance and universality stems from the central limit
theorem, that states that the normal distribution arises from
averaging a sufficiently large number of random variables, regardless
of which distribution rules their fluctuations on an individual basis
(provided some properties to which we return shortly).  While the
normal (ergodic) type of fluctuations describes much of the physical
word, it can fail dramatically for highly complex systems.  A
different class of fluctuations, for which the central limit theorem
does not hold, provides a new paradigm. They are known as power law or
fat-tail type of distributions,\cite{coles_book01a} the most famous
and widespread type of which is, in physics, the Lorentzian. The main
property of such fluctuations is that they are scalefree, that is, the
fluctuator can wander arbitrarily far from its most likely position,
in contrast with a normal distribution where $5\sigma$ deviations are
so unlikely that they are regarded as ``discovery'' of a new physics
(the concept of ``standard deviation'' is rooted in the normal
distribution and has no counterpart for scalefree fluctuators).  When
it is combined with impact of the outliers, scalefree fluctuations
give rise to new notions such as kurtosis-risk,
extreme-risk\cite{garrett08a} and ``black
swans''\cite{taleb_book07a}---highly unlikely events that can totally
alter the trajectory or response of the system.\footnote{As in our
  case the phenomenon is reproducible, it would be more accurately
  referred to as a ``gray swan'' by proponents of extreme outliers
  theories.}

In the following, we assume that our elementary quantum optical system
is embedded in such an environment that has scalefree fluctuations.
We will assume a Lorentzian type of fluctuation, but qualitatively
similar results would follow from other types of scalefree
fluctuations.\footnote{The formalism we have presented can be used for
  any type of fluctuations, including those of the Gaussian type, the
  case of most likely interest. Applied to the system reported in this
  letter, Gaussian fluctuations result in a broadening only of the PL
  line and no FIL.} We are thus left to express the rates
$\phi_{R\rightarrow R'}$ which enforce the steady state solution of
$\partial_t \mathcal{P}_R=-\sum_{R'}\phi_{R\rightarrow
  R'}\mathcal{P}_R+\sum_{R'}\phi_{R'\rightarrow R}\mathcal{P}_{R'}$ to
be proportional to the Lorentzian distribution
$\mathcal{P}_R=\frac{1}{\mathcal{N}}\frac{1}{\pi}\frac{\gamma_F/2}{(\gamma_F/2)^2+(R\delta)^2}$,
with $\sum_R \mathcal{P}_R =1$.  $\mathcal{N}$ is the normalization
constant that removes the units (and can be simplified in the
continuum limit) $\mathcal{N}=\sum_R
\frac{1}{\pi}\frac{\gamma_F/2}{(\gamma_F/2)^2+(R\delta)^2}\approx
\frac{1}{\delta}\int dx
\frac{1}{\pi}\frac{\gamma_F/2}{(\gamma_F/2)^2+x^2} =
\frac{1}{\delta}$.  This is obtained with rates of the form:
\begin{equation}
  \label{eq:ThuNov24113218CET2011}
  \phi_{R\rightarrow R'}=\phi\mathcal{P}_{R'} \approx \frac{\phi}{\delta}\frac{1}{\pi}\frac{\gamma_F/2}{(\gamma_F/2)^2+(R'\delta)^2}\,,
\end{equation}
with $\phi$ the speed of fluctuations. They only depend on the final
state. The transition towards the central point has the largest rate,
being the most likely one no matter from which point the transition is
initiated.

We can now solve these equations numerically. A typical result is
presented in \ref{fig:FriJul20142005CEST2012}(c) and, clearly,
reproduces the same phenomenology observed in our experiments; namely
the unexpected emission line, denoted ``FIL'' for Fluctuation Induced
Luminescence.  This new line arises from the fluctuations as sketched
in \ref{fig:FriJul20142005CEST2012}(d): when the system is far
detuned, it infrequently wanders into resonance with the laser by the
very nature of the scalefree fluctuation that allows such giant
deviations. The photo-generated electron-hole pair will then relax
into the s-shell before emitting light. Depending on the speed of the
fluctuations, the emitted photon can either have the frequency of the
most likely s-shell energy when the fluctuations are faster than the
recombination time, or it can have a frequency which is detuned from
the most likely s-shell energy when the fluctuations are slower than
the recombination time.  The strong absorption gain which occurs there
compensates for the combined scarcity and brevity of these
events. This is depicted in \ref{fig:FriJul20142005CEST2012}(d) with
Lorentzian fluctuations of the levels as a function of time with three
highlighted emission processes: i) from an extreme outlier which emits
far from the most likely s-state energy, producing no notable
luminescence as such events are rare and emit weakly, ii) from a state
close to the most likely s-state energy, producing PL as a result of
the many times this weak-emission configuration is realized, and iii)
from an extreme outlier which, by chance, brings the system into
resonance, producing detectable luminescence from the strong emission
compensating for the rarity of such events. Our system, therefore,
implements ``black-swans'' in the solid state: extremely rare
occurrences that have a huge impact and result in a qualitative change
in the system.

One can study from numerical solutions the properties of the FIL peak.
Depending on parameters (such as when $\gamma_{21}<\gamma_F$ and
$\phi$ is small), the FIL peak is sharper and brighter than the
PL. Interestingly, it is found that the speed of fluctuations is an
important ingredient that determines the form of the detuning
dependent FIL and PL emission. Faster fluctuations decrease the
intensity of FIL leaving its linewidth essentially unchanged.  In the
limit of slow fluctuations, numerical solutions converge towards the
analytical expression obtained by integrating the photoluminescence
spectrum over its fluctuations: $\tilde
S(\omega,\Delta)=\int_{-\infty}^{\infty}S(\omega,\delta)\frac{1}{\pi}\frac{(\gamma_F/2)}{(\gamma_F/2)^2+(\delta-\Delta)^2}\,d\delta$. Plugging
Eq.~(\ref{eq:FriNov4124436CET2011}) in this expression, one can
compute the slow Lorentzian fluctuations PL spectrum in the linear
regime in a closed form: it is the sum of two peaks, with their
Lorentzian and dispersive parts given by:
\begin{subequations}
  \label{eq:MonNov21115227CET2011}
  \begin{align}
    &S(\omega,\Delta)=\frac{\Omega^2/\gamma_{10}}{\Big[ (\frac{\gamma_F-\gamma_{21}}{2})^2+\Delta^2\Big]\Big[ (\frac{\gamma_F+\gamma_{21}}{2})^2+\Delta^2\Big]}\sum_{p=\mathrm{FIL\,,PL}}\frac{1}{\pi}\Big[\frac{L_p\gamma_p/2-K_p(\omega-\omega_p)}{(\gamma_p/2)^2+(\omega-\omega_p)^2}\Big]\,,\\
    &\omega_\mathrm{FIL}=0\,,\quad\gamma_\mathrm{FIL}=\gamma_{10}+\gamma_{21} \,,\quad L_\mathrm{FIL}=\gamma_F\Big[ \frac{\gamma_F^2-\gamma_{21}^2}{4}+\Delta^2\Big]\,,\quad K_\mathrm{FIL}=-\gamma_F\gamma_{21}\Delta\,,\\
    &\omega_\mathrm{PL}=\Delta\,,\quad\gamma_\mathrm{PL}=\gamma_{10}+\gamma_F
    \,,\quad L_\mathrm{PL}=\gamma_{21}\Big[
    \frac{\gamma_{21}^2-\gamma_F^2}{4}+\Delta^2\Big]\,,\quad
    K_\mathrm{PL}=\gamma_F\gamma_{21}\Delta\,.
\end{align}
\end{subequations}
This fully describes the properties of the peaks depending on the
system parameters and the amplitude of fluctuations when they are
slow. This shows in particular that the peak which is sharper is also
the one that is the more intense. Not all these attributes are
conserved in the exact numerical solution when the speed of
fluctuations is not very slow, but at least the qualitative result is
spelt out and easily understood in this limiting case.

\begin{figure}[th]
  \includegraphics[width=.9\linewidth]{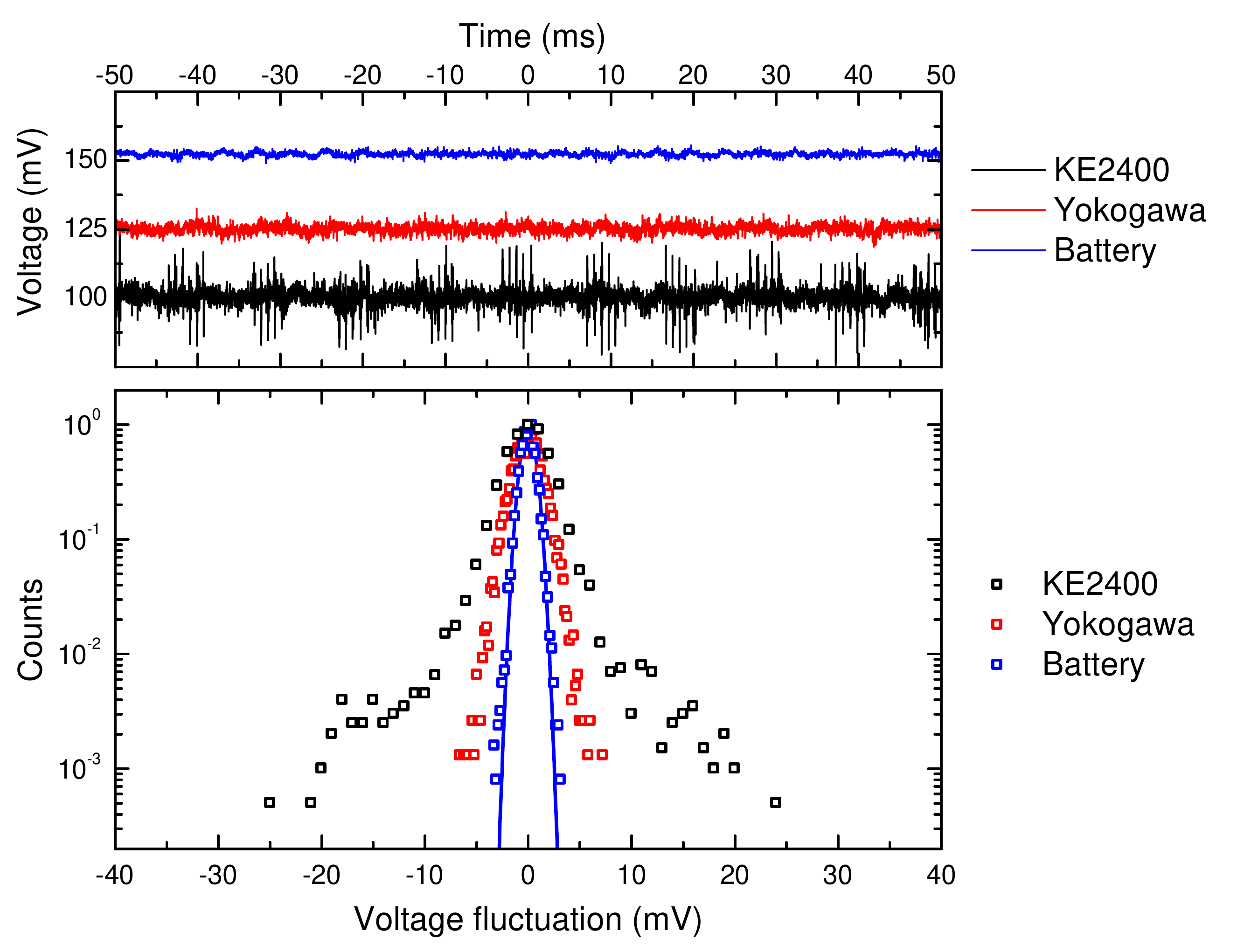}
  \caption{Fluctuations due to the voltage source. The upper row shows
    a trace of the applied voltage over a 10ms window, for three
    sources: (a) a standard Keithley KE2400 source, (b) a Yukogawa
    source and (c) a simple battery. Beside larger amplitudes of
    fluctuations the Keithley and Yukogawa have fluctuations to all
    orders, that is, infrequent but large deviation from the applied
    voltage. These are the cause of FIL.}
  \label{fig:FriJul20143318CEST2012}
\end{figure}

The FIL peak in the experiment is observed when the excitation laser
is detuned by $\Delta\leq0.5$meV from the excited orbital state
resonance. At the largest detunings where FIL is still clearly
observable, defined as being when its intensity is at least 1/4 of
that of the PL, the system would be more than 20 standard deviations
away from the median if it would follow a normal distribution. The
probability for the system to fluctuate into resonance according to
the Lorentzian distribution is approximately one in a
million. However, since the quantum dynamics occurs over picosecond
timescales, and the spectra are recorded with integration times
extending over several minutes, this still leaves sufficient
opportunities for these extremely rare events to occur and give rise
to the FIL peak in the spectrum.  They are observable as compared to
off-resonant cases thanks to their huge impact.  The same probability
is completely negligible according to the Gaussian distribution and
would not be once realized, even if running the experiment over
several lifetimes of the universe.

\begin{figure}[th]
  \includegraphics[width=.9\linewidth]{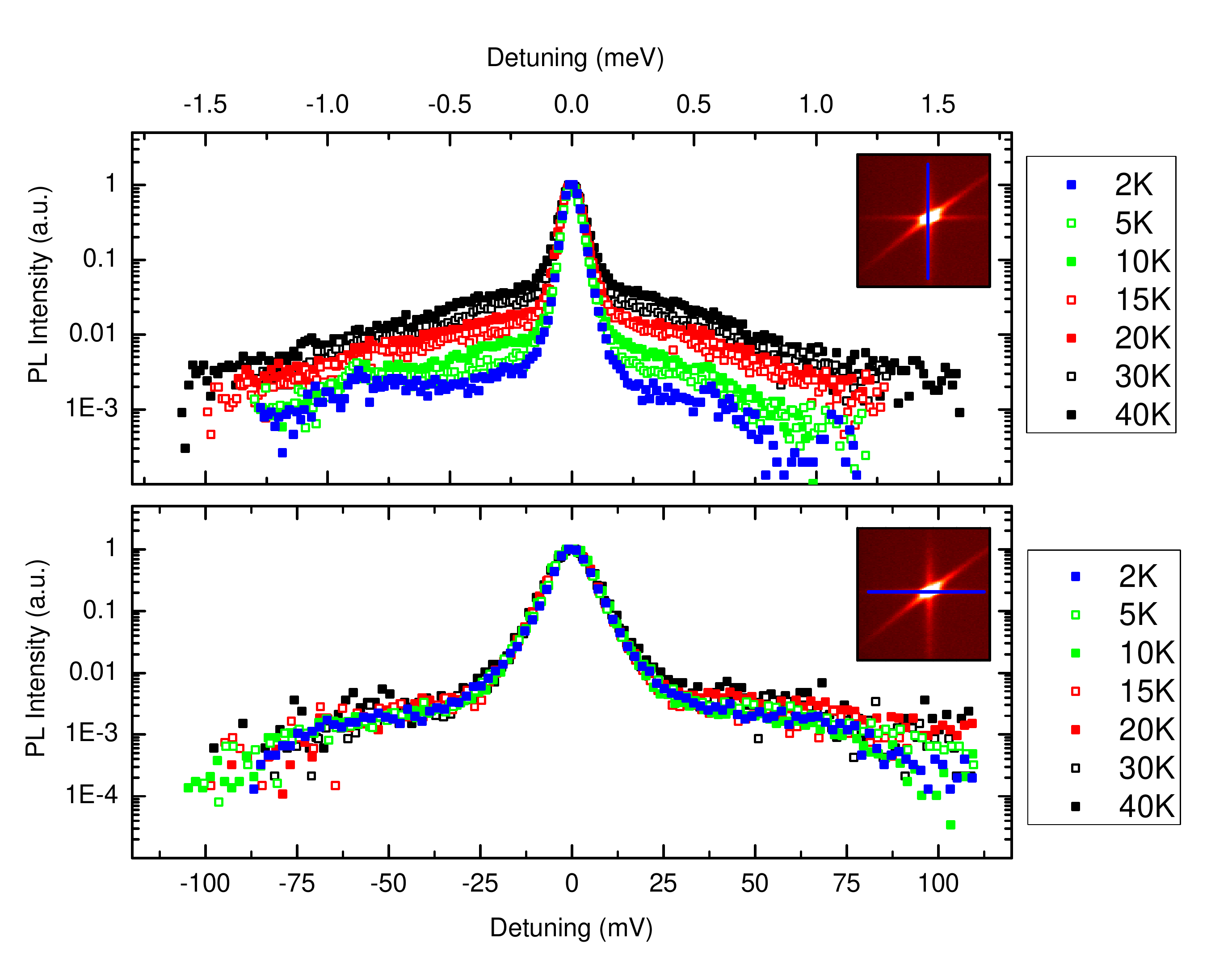}
  \caption{Temperature series for the cross-section of (a)
    luminescence at resonance and (c) FIL. While the former exhibits a
    clear and characteristic temperature dependence, the FIL is
    temperature independent, pointing at an extrinsic origin for the
    scalefree fluctuations causing it.}
  \label{fig:FriJul20142832CEST2012}
\end{figure}

Fingerprints of non-ergodic fluctuations are actually commonplace in
the emission from individual solid state emitters, including
flurophores such as molecules,\cite{hoogenboom05a} proteins and
polymers,\cite{dickson97a,vandenbout97a} semiconductor quantum
dots,\cite{nirmal96a,robinson00a} nanorods\cite{wang06a} and
nanowires. Ubiquitous phenomenology such as fluorescence
intermittency\cite{frantsuzov08a,nirmal96a} and spectral
wandering\cite{empedocles96a} are observed arising from coupling to
their local environment. In our system, the source of the fluctuation
was traced to be fluctuations of the programmable voltage source in
which electronics stabilize the output voltage. Scalefree fluctuations
are of an entirely counterintuitive character and the cause of many
unpredictable and important events in complex environments, such as
catastrophes on financial markets\cite{mandelbrot_book04a} or sudden
large-scale changes in meteorological systems.\cite{garrett08a} In our
case, tracing their origin to the electronics in the setup was a
difficult and time-consuming task since such fluctuations are
typically invisible to the spectral noise of the devices. Furthermore,
such unlikely events do not appear in the technical description of the
source provided by the manufacturer. It is not established, for
instance, that they are exactly of the Lorentzian type, but their
fat-tail property is mandatory to understand our experimental
observations. In most experiments, such extremely rare fluctuations of
voltage have no impact whatsoever. It requires the black-swan
character, only met in particular configurations such as ours,
stemming from the extremely large gain of the QD absorption driven
quasi resonantly that make these rare events impactful. In our QDs and
QDMs, their effect was remarkable since they produced a new line in
the emission spectrum. The FIL shown in
\ref{fig:FriJul20142005CEST2012} was obtained with a Keithley 2400
voltage source. We have repeated the experiment using a Yokogawa
voltage source, and observed that the FIL peak reduced significantly
in intensity, although it did not disappear completely. We have been
able to remove the scalefree fluctuations entirely by using a simple
battery-driven voltage follower circuit. The voltage traces measured
for these three sources are shown in
\ref{fig:FriJul20143318CEST2012}. The KE2400 series clearly exhibits
random fluctuations with large deviations from the median, resulting
in a clear characteristic fat-tail profile in its distribution (see
the lower panel in \ref{fig:FriJul20143318CEST2012}). The Yokogawa
source, on the other hand, appears much more Gaussian-like with no
observable large departures. The fact that a FIL is still observed
shows that it, too, features extreme outliers with scalefree
properties, but that we are unable to record them in the time windows
over which the voltage was monitored. The difficulty to detect and
characterize power-law distributions is a well-known problem of
statistical analysis.\cite{clauset09a} By turning these elusive
outliers into black swans, our system therefore implements a way to
probe and characterize such fluctuations. It could allow, for
instance, to quantify the range over which the power law holds, one of
the most difficult practical problem in the field.  Such fluctuations
could also be intrinsic, for instance from trapping of a carrier in a
nearby defect, with fluctuations in the induced electric field
correlated with the time spent by the carrier in the trap, which being
memoryless in time would result in Lorentzian fluctuations.  In our
case, however, we can rule out such possibilities from a temperature
series for the conventional and fluctuation-induced luminescence,
both shown in \ref{fig:FriJul20142832CEST2012}. While the conventional
PL exhibits the characteristic phonon-sidebands with their hallmark
asymmetry at low temperature,\cite{muljarov05a} the FIL is temperature
independent, showing that it has no connection with the intrinsic
fluctuation spectrum or dynamic of the system.

In conclusion, we have presented a general theory of fluctuations in a
dissipative quantum optical system and applied it to the case of a
resonantly driven quantum dot in the presence of scalefree
fluctuators. We have shown that in a system exhibiting strong gain,
the combination of rare and impactful events may result in drastic and
qualitative changes in the system, known in the stochastic literature
as ``black swan events''.  In our case, a new sharp line pinned at the
laser excitation is produced. This type of luminescence has never been
reported before to the best of our knowledge, and further
investigations are needed to study whether they might prove useful for
applications and/or characterization.  The system we have presented
could be used as a sensitive probe of its fluctuating environment and
a powerful gauge of power-law tails. Although such fluctuations could
be intrinsic to the system, we have reported here an experimental
occurrence that was put forward, most unexpectedly, by the voltage
source apparatus, and an enduring puzzle was put to rest with a simple
battery.

% \begin{acknowledgement}
% \end{acknowledgement}

% \begin{suppinfo}

% This will usually read something like: ``Experimental procedures and
% characterization data for all new compounds. The class will
% automatically add a sentence pointing to the information on-line:

% \end{suppinfo}

\bibliography{Sci,books,FIL}

%% graphical table of contents.
%%%%%%%%%%%%%%%%%%%%%%%%%%%%%%%%%%%%%%%%%%%%%%%%%%%%%%%%%%%%%%%%%%%%%

% \begin{tocentry}

% Some journals require a graphical entry for the Table of Contents.
% This should be laid out ``print ready'' so that the sizing of the

% Inside the \texttt{tocentry} environment, the font used is Helvetica
% 8\,pt, as required by \emph{Journal of the American Chemical

% The surrounding frame is 9\,cm by 3.5\,cm, which is the maximum
% permitted for  \emph{Journal of the American Chemical Society}
% graphical table of content entries. The box will not resize if the
% content is too big: instead it will overflow the edge of the box.

% This box and the associated title will always be printed on a
% separate page from the rest of the document. It is best to place the
% graphical TOC entry as the last item in the draft so that the page
% break is not a problem.

% \end{tocentry}

This work is funded by the Deutsche Forschungsgemeinschaft via Grant
No. SFB-631 and the excellence cluster Nanosystems Initiative Munich
and the European Union via SOLID (FP7-248629) and S3Nano
(FP7-289795). FPL acknowledges support from the Marie Curie IEF
``SQOD'', EdV from the Alexander von Humboldt Foundation. G.A. also
thanks the Technische Universit\"at M\"unchen Institute for Advanced
Study for support.

\end{document}